\documentstyle[sprocl]{article}

\def\jou#1#2#3#4{{#1} {\bf #2}, #3 (19#4)}

\def\NC{\em Nuovo Cimento}
\def\NP{\em Nucl. Phys.}
\def\PL{\em Phys. Lett.}
\def\PRL{\em Phys. Rev. Lett.}
\def\PR{\em Phys. Rev.}
\def\RMP{\em Rev. Mod. Phys.}
\def\ZP{\em Z. Phys.}
\def\tiz{SU(4)_{PS}\otimes SU(2)_L\otimes SU(2)_R}
\def\lu{SU(3)_c\otimes SU(2)_L\otimes SU(2)_R\otimes U(1)_{B-L}}
\def\sm{SU(3)_c\otimes SU(2)_L\otimes U(1)_Y}
\def\SM{G_{SM}}
\def\ul{SU(3)_c\otimes U(1)_Q}
\def\rw{\rightarrow}
\def\lrw{\longrightarrow}
\def\pd{{p\rw e^+\pi^0}}
\def\al{\alpha}
\def\gp{\Phi}
\def\ru1{\rule[-0.4truecm]{0mm}{1truecm}}
\newcommand{\gapproxeq}{\lower .7ex\hbox{$\;\stackrel{\textstyle >}{\sim}\;$}}
\newcommand{\lapproxeq}{\lower .7ex\hbox{$\;\stackrel{\textstyle <}{\sim}\;$}}
\def\be{\begin{equation}}
\def\ee{\end{equation}}
\def\bea{\begin{eqnarray}}
\def\eea{\end{eqnarray}}
\def\ba{\begin{array}}
\def\ea{\end{array}}

\begin{document}

\title{\hfill $\mbox{\small{
$\stackrel{\rm\textstyle DSF-47/97}
{\rm\textstyle hep-ph/9709416~\quad\quad}$}}$ \\[1truecm]
SO(10) Unified Theories and Cosmology \footnote{presented by F. 
Buccella at NANP97, Dubna, July 1997.}}

\author{ F. Buccella, G. Mangano, O. Pisanti, and L. Rosa, }

\address{Dipartimento di Scienze Fisiche, Universit\'a di Napoli ``Federico
II'', \\ 
Pad. 19 Mostra d'Oltremare, I-80125 Napoli, Italy \\
and INFN, Sezione di Napoli, \\
Pad. 20 Mostra d'Oltremare, I-80125 Napoli, Italy.}

\maketitle\abstracts{
We review the status of a class of gauge unified models based on SO(10)
group. After a pedagogical introduction to SO(10) gauge theories, we
discuss the main phenomenological implications of these models. The upper
limit on proton lifetime are obtained and the prediction for neutrino
masses are compared with the astrophysical and cosmological constraints
coming from solar neutrino data and dark matter problem. Possible scenarios
for the production of the baryon asymmetry of the universe required by
primordial nucleosynthesis are also discussed.} 

{\it This paper is dedicated to the memory of Prof. Roberto Stroffolini,
whose friendship and deep humanity enriched the authors. We will not forget
his rigorous approach to physics and his invaluable effort and enthusiasm
in over forty years of teaching activity.} 

\section{Introduction}

We will review the present status of a class of non SUSY Grand Unified
Theories (GUT's) based on the simple gauge group SO(10). 

The unification of electromagnetic and weak interactions, achieved in the
framework of Glashow-Salam-Weinberg $SU(2)_L\otimes U(1)_Y$ model
\cite{glas}, has been experimentally tested with a remarkable precision
\cite{2}. Similarly, Quantum Chromodynamics, based on $SU(3)_c$ gauge
group, is well established as the theory of strong interactions, though
infrared slavery prevents us from a clear understanding of low energy
phenomena and still the mechanism of confinement remains obscure. The
Standard Model (SM) of elementary interactions, the $\sm$ gauge theory, is
without any doubt already a piece of history of science. GUT's represent,
along the same ideological line, a further effort towards a simplified
picture of the elementary particle world. It is worth reminding that the
idea that all interactions can be described by a simple group gauge theory
at very high energies, more than a theoretical prejudice, relies upon the
fact that the three SM running coupling constants converge towards a common
value as the energy scale increases, suggesting that at some scale $M$,
interactions may undergo a phase transition to a different behaviour
characterized by a larger gauge symmetry. 

One may wonder how such a new scenario could be ever tested experimentally
if the scale $M$, as suggested by many arguments, is so high ($10^{15}\div
10^{16}\,GeV$) to be far away from the detection possibility of present and
future accelerators. However, its presence would indirectly affect low
energy physics by very tiny effects, proportional to some negative power of
$M$. The most famous example is the prediction of proton instability, which
actually is a peculiar signature of GUT's. In this case the decay rates for
typical channels as $\pd$ or $\pi^+\bar\nu_\mu$ are expected very small,
being proportional to $M^{-4}$. Present and future experiments as
Super-Kamiokande \cite{tots} or ICARUS \cite{rubb} will test the
interesting region for proton decay channel rates of $10^{32}\div 10^{33}$
years. In this respect, the {\it minimal} GUT, based on SU(5) group
\cite{gegl}, is already at variance, in its minimal version, with the
precise measurements of the SM coupling constants at the $Z^\circ$ mass
scale and the present bound on proton lifetime \cite{ambu}. 

In SO(10) GUT the interplay between low energy phenomena and large scales
also may show up in neutrino physics. On the basis of the {\it see-saw}
mechanism \cite{gera}, (almost) left-handed neutrinos acquire masses of the
order $m^2/M$, with $m$ of the order of the up quark mass of the same
generations. As we will discuss, if $M$ represents an intermediate symmetry
scale of the order of $10^{11}\,GeV$, this would predict $m_{\nu_e}\ll
m_{\nu_\mu}\sim 10^{-3}\,eV$, $m_{\nu_\tau}\sim 10\,eV$. The $\mu$ neutrino
mass is actually in the range to explain, in the framework of the MSW
mechanism \cite{wolf}, the solar neutrino flux deficit observed by many
experiments \cite{gall}. Masses for $\nu_\tau$ larger than $1\,eV$ will be
instead observed by CHORUS and NOMAD Collaborations \cite{chor}, and would
render $\nu_\tau$ the main contribution to the {\it hot} component of dark
matter (DM). 

The two tests for SO(10) GUT's just mentioned demonstrate how important the
interplay between particle physics, astrophysics and cosmology became in
the last ten years, mainly due to an astonishing increase in precision of
astrophysical measurements. This fact is particularly relevant, since
effects which are instead proportional to $M$, i.e. which took place in the
very early universe, can be tested by looking at the way they influenced
the subsequent evolution of the universe. The baryon asymmetry, constrained
by observation on primordial light nuclei abundances \cite{oliv}, the
production of topological defects as monopoles or cosmic strings, finally
the density perturbations caused by an inflationary epoch, provide a
coherent set of severe constraints on GUT's. 

We hope we succeeded in communicating our strong feeling that, during next
decade, observations of the universe will tell us many things about GUT's,
either confirming their role at high energy scales or ruling them out. This
report is organized as follows: in section 2 we give a short pedagogical
introduction to SO(10) GUT's. Readers who are familiar with the subject can
skip it and directly go to section 3, where we discuss a class of SO(10)
models with $\tiz$ or $\lu$ intermediate symmetry and which have been
extensively studied in the past ten years \cite{bucc}. Section 4 is devoted
to a discussion of many phenomenological implications of SO(10) GUT's and,
in particular, of the models described in section 3. As far as the
conclusions, we advice the reader to read again this Introduction and
section 5. 

\section{An introduction to SO(10) GUT's}

We will here shortly review the main features of SO(10) GUT's. More
detailed discussions can be found in \cite{geor}. 

A good starting point is perhaps to recall the classification of
left-handed fermions in the SM; under $\sm$ we have 
\be
\psi_L\sim (1,1,1)\oplus \left(3,2,\frac{1}{6}\right)\oplus \left(\bar
3,1,-\frac{2}{3}\right) \oplus \left(1,2,-\frac{1}{2}\right) \oplus
\left(\bar 3,1,\frac{1}{3}\right). 
\ee
One may wonder if a simplification of this picture is possible, also
allowing for a natural explanation of the electric charge quantization, by
embedding the SM gauge group in a larger one G. If G is chosen to be a
simple group, then the three independent gauge coupling would merge in one
only. The first model realizing all this was proposed over twenty years ago
by Georgi and Glashow \cite{gegl}, based on the group SU(5)
\footnote{Actually SU(5) is the smallest group whose algebra contains
$su(3) \otimes su(2) \otimes u(1)$ as a maximal subalgebra.}. In this case
the number of fermion representations is reduced to two only, 
\be
\psi_L\sim 10\oplus \bar 5.
\ee
The choice of 24 and $5\oplus \bar 5$ dimensional representations for the
Higgs bosons gives the desired symmetry breaking pattern, 
\be
SU(5)\buildrel 24\over\lrw \sm\buildrel 5\oplus\bar 5\over\lrw\ul,
\ee
and the right quantization of hypercharge Y and electric charge Q
\cite{abbu}. There is also a beautiful prediction on fermion masses, due to
the fact that, for each generation, down antiquark and lepton doublet are
contained in one representation, 
\be
m_b\sim 3 m_\tau,
\ee
once the masses are evolved down to low scales from the SU(5) unification
scale \cite{buel}. 

A glance to the vector gauge bosons, contained in the adjoint 24
representation, 
\be
A\sim 24\sim (8,1,0)\oplus (1,3,0)\oplus (1,1,0)\oplus
\left(3,2,-\frac{5}{6}\right)\oplus \left(\bar 3,2,\frac{5}{6}\right), 
\ee
shows the presence of usual SM vector bosons as well as very massive {\it
leptoquarks}, which allow for nucleon instability via processes like $\pd$,
whose rate is of the order of $\al^2 m_p^5 M^{-4}$. The present enormous
lower limit on proton lifetime for this channel, $\tau_\pd \gapproxeq
10^{33}$ years \cite{gaje}, therefore would require the SU(5) unification
scale $M$ to be larger than $10^{15}\,GeV$. 

Why abandon minimal SU(5) \footnote{It is worth pointing out that more
complicated choices for the Higgs boson representation  or SUSY SU(5) are
in agreement with all available data and the following considerations do
not apply.}? Since SU(5) directly breaks down to SM, one expects that the
three couplings $\al_s$,  $\al_2$ and $\al_Y$ should meet at the
unification scale $M_{SU(5)}$. However, using the measurements of $\al_i$
at the $M_Z$ scale \cite{pdg} and assuming that only customary SM particles
contribute to the renormalization group equations (RGE) \cite{ambu}, the
three couplings meet at three different points and only the scale at which
$\al_2=\al_s$ is large enough ($\gapproxeq 10^{16}\,GeV$) to be in
agreement with the lower limit on proton lifetime. If this experimental
evidence rules out minimal SU(5), it suggests on the other hand that
unification may proceed through an intermediate symmetry stage. It has been
observed, for example in the sixth reference of \cite{bucc}, that, if
hypercharge receives a contribution from a generator of a non abelian
group, as it is the case for SO(10) GUT's, this would reconcile the
experimental data with a GUT scheme. 

SO(10) GUT theories were proposed many years ago \cite{geor} on the basis of
completely independent motivations: 
\begin{enumerate}
\item For each generation, all left-handed fermions are classified in only
one irreducible representation, the 16-dimensional spinorial
representation. Under SU(5) it decompose as $10\oplus \bar{5}\oplus 1$,
where the additional singlet, with respect to the SU(5) case, has the
quantum numbers of $\nu^c_L$. The presence of this state, sterile under the
SM and SU(5) actions, is a consequence of the possibility to define in
SO(10) a charge conjugation operator  $\cal C$ (which is not the usual
Dirac one) which is a linear combination of the algebra's generators. Under
$\cal C$, the left-handed weak interacting neutrino state transforms into
$\nu_L\buildrel {\cal C}\over\lrw \nu^c_L$. 
\item Models based on SO(10) gauge group are naturally anomaly free. What
can be regarded as a lucky circumstance in SU(5) model, because of their
exact compensations for the 10 and $\bar{5}$ representations, is instead a
general feature of orthogonal groups, with the only exception of SO(6). 
\item There is an intriguing decomposition of the 16 under the Pati-Salam
group $\tiz$ \cite{pasa}, 
\be
16=(4,2,1)\oplus(\bar{4},1,2),
\ee
which displays the quark-lepton universality of weak interactions.
\end{enumerate}

Baryon number violation and proton instability is a feature of SO(10) GUT's
as well. Among the gauge vector bosons, classified in the 45-dimensional
representation, there are even more {\it leptoquark} states than in SU(5)
case which can mediate nucleon decay. In particular, decomposing the SO(10)
adjoint representation under $\lu$ (this subgroup will play a relevant role
as intermediate symmetry stage in the following) we have 
\bea
45 &=& (8,1,1,0)\oplus (1,3,1,0)\oplus (1,1,3,0)\oplus (1,1,1,0)\oplus
\left(3,1,1,\frac{4}{3}\right) \nonumber \\ 
&& \oplus \left(\bar{3},1,1,-\frac{4}{3}\right)\oplus
\left(3,2,2,-\frac{2}{3}\right)\oplus \left(\bar{3},2,2,\frac{2}{3}\right).
\eea
The fact that the baryon and lepton number difference B-L is gauged in
SO(10) GUT's, and is eventually spontaneously broken at low scales, has
important consequences for the production of a baryon asymmetry in the
universe. We will come back to this point in section 4. It is also worth
noticing  that SO(10) embeds $SU(2)_L\otimes SU(2)_R$. The scales at which
the two groups break down, say $M_L$ and $M_R$, are however quite distinct
since $M_L$ is of the order of the electroweak scale while $M_R$ is
expected to be very large ($\sim10^{11}~GeV$). Actually baryon number
generation is also a way to probe the difference $(M_R-M_L)/M_X$, where
$M_X$ is the SO(10) breaking scale. 

It is also remarkable the way weak hypercharge can be written in terms of
right isospin $T^3_R$ and B-L, 
\be
Y=T_R^3+\frac{B-L}{2}.
\ee
When $SU(2)_R$ gauge symmetry is restored, weak hypercharge coupling
therefore receives a contribution in scale evolution by a non abelian
factor, which results into a change of the corresponding $\beta$ function
from positive to negative. This may shift the $\al_Y-\al_2$ intersection
point of SU(5) prediction up to the larger scale when also $\al_2$ and
$\al_3$ meet. 

Fermion masses in SO(10) GUT's may be produced via the usual symmetry
breaking mechanism and Yukawa couplings to Higgs bosons of the form 
\be
10_H\cdot(16_F\otimes16_F)_{10},~126_H\cdot(16_F\otimes16_F)_{126},~
120_H\cdot(16_F\otimes16_F)_{120}.
\ee
All fermion Dirac masses are expected to be generated at the very last
stage, when SM breaks down to $\ul$. The presence of both $\nu_L$ and
$\nu^c_L$ in the 16 representation however, along with a Dirac term, 
\be
m^D\nu^{cT}_L\sigma_2\nu_L,
\ee
allows for a Majorana mass,
\be
m^M=\nu^{cT}_L\sigma_2\nu^c_L,
\ee
which appears when both $SU(2)_R$ and $U(1)_{B-L}$ are spontaneously broken
\footnote{A similar Majorana term could in principle be added for $\nu_L$
states but the addition of a $SU(2)_L$ Higgs triplet of high mass would
change the ratio $M_Z/M_W$.}. The neutrino mass matrix therefore takes the
form, up to radiative corrections, 
\be
m_\nu=\left(\begin{array}{cc}
0 & m^D \\  m^D & m^M
\end{array} \right).
\ee
Because the scale $M_R$ at which $SU(2)_R$ is broken is much higher than
$M_L$, it follows that the two mass eigenvalues are approximately equal to
$m^M$ and $(m^D)^2/m_M\ll m^M$. Thus, this {\it see-saw} mechanism for
neutrino masses \cite{gera} predicts an (almost) right-handed very heavy
neutrino and a very light (almost) left-handed one, much lighter, for a
factor $m^D/m^M\sim M_L/M_R$, than the charged lepton or quarks of the same
generation. This beautiful prediction of SO(10) GUT's may explain why weak
interacting neutrinos are expected very light (though till now they could
be well massless states!). If $M_R$ is of the order of $10^{11}~GeV$, as
noticed in \cite{bucc,mopa,bash}, neutrino masses may be in the right range
to explain the {\it solar neutrino problem} in the MSW scheme and to
account for the {\it hot} component of DM (see section 4). 

We close this short summary of SO(10) GUT's with some remarks on the
symmetry breaking pattern. In general the pattern from SO(10) down to the
SM gauge group depends on the Higgs boson representations which are
considered. There is in fact quite a large variety of models, leading to
different results for the unification scale. A common feature of all these
model is, however, that the symmetry breaking takes place via an
intermediate stage, with group symmetry $G' \subset SO(10)$, 
{
\small
\be
SO(10)\buildrel M_X\over\lrw G'\buildrel M_R\over\lrw\SM\buildrel
M_{EW}\over \lrw\ul.
\ee
}
This result holds for all models based on Higgs chosen in the low
dimensional representations (10, 16, 45, 54, 120, 126, 210) \footnote{The
only exception is the 144-dimensional representation \cite{pisa}.}, since
in all these cases the components invariant under $\sm$ have little groups
larger than the SM group. We have indicated the second symmetry breaking
scale with $M_R$, since it typically corresponds with the breaking of
$SU(2)_R$, though this is not always the case. 

\section{A Class of SO(10) Models}

In this section we discuss in more details a class of models of SO(10)
GUT's with $\tiz$ or $\lu$ intermediate symmetry group \cite{bucc}. In
table 1 are reported the four possible intermediate groups $G'$ along with
the Higgs representation used to break SO(10). In all cases the second
symmetry breaking at the scale $M_R$ is realized using the
$126\oplus\overline{126}$ bispinorial representations. Actually this could
be achieved using spinorial representation 16 as well, but this would
result in too small Majorana masses for right-handed neutrinos (and via the
{\it see-saw} mechanism, too large masses for left-handed ones)
\cite{witt}. 

\begin{table}[t]
{\scriptsize
\begin{center}
\begin{tabular}{|c|c|c|c|} \hline
&&& \\
& $G'$ & Higgs direction & Repr. \\ 
&&& \\ \hline
&&& \\
A & $\tiz\times D$ & $\omega_L=\frac{2\left(\omega_{11}+ \ldots+
\omega_{66} \right)- 3\left(\omega_{77}+ \ldots
\omega_{00}\right)}{\sqrt{60}}$ & 54 \\ 
&&& \\ \hline
&&& \\
B & $\lu\times D$ & $\gp_L=\frac{\gp_{1234}+ \gp_{1256}+
\gp_{3456}}{\sqrt{3}}$ & 210 \\ 
&&& \\ \hline
&&& \\
C & $\tiz$ & $\gp_T=\gp_{7890}$ & 210 \\
&&& \\ \hline
&&& \\
D & $\lu$ & $\gp(\theta)=\cos\theta\gp_L+ \sin\theta\gp_T$ & 210 \\ 
&&& \\ \hline
\end{tabular}
\end{center}}
\caption{Four possible intermediate symmetry groups in SO(10) GUT's. D is
the left-right discrete symmetry. $\omega_{ab}$ is a second-rank traceless
symmetric tensor; $\gp_{abcd}$ is a fourth-rank antisymmetric tensor, and
the indices 1...6 correspond to $SO(6)\sim SU(4)_{PS}$, whereas 7...0
correspond to $SO(4)\sim SU(2)_{L}\otimes SU(2)_{R}$.} 
\end{table}

Once the correct spontaneous symmetry breaking pattern is realized, the
main goal is to obtain informations on the unification scales $M_X$ and
$M_R$. This is done by evolving the SM coupling constants, experimentally
known at the $M_Z$ scale \cite{pdg}, 
\begin{eqnarray}
\sin^2{(\theta_W)} &=& 0.2315 \pm 0.0002,\nonumber \\
\al_s &=&0.120 \pm 0.005, \\
\al_{em} &=& (127.9 \pm 0.09)^{-1}, \nonumber
\end{eqnarray}
with the energy scale using the RGE,
\be
\mu\frac{d}{d\mu}\al_i(\mu)=\beta_i\left( \al_i(\mu) \right).
\ee
The scales $M_X$ and $M_R$ are then obtained by requiring that SO(10) or
$G'$ symmetries are restored. The main problem in this procedure is that
there is usually a huge number of Higgs scalars which contributes to RGE as
soon as the scale becomes larger than their mass.  It is customary to adopt
a simplifying assumption, the Extended Survival Hypothesis (ESH)
\cite{bamo}, i.e. to consider in RGE only those scalars which are required
to drive symmetry breaking at $M_R$ and at the electroweak scale. In this
case one gets \cite{bucc} (the results are updated for the new values on SM
gauge couplings)

\smallskip

$\!\!\!\!\!\!\!\!\!\!\!\!\!$
\begin{tabular}{|c|c|c|c|c|c|}
\hline\ru1
& $M_X/10^{15}\,GeV$ & $M_R/10^{11}\,GeV$ &
& $M_X/10^{15}\,GeV$ & $M_R/10^{11}\,GeV$ \\
\hline\ru1
A & 0.6 & 460 & C & 4.7 & 2.8 \\
\hline\ru1
B & 1.6 & 0.7 & D & 9.5 & 0.067 \\
\hline  
\end{tabular}

\smallskip

The phenomenological implications of these results will be discussed in
section 4. We only notice here that models with the left-right $D$ symmetry
\cite{kush} give smaller values for $M_X$ and so shorter proton lifetime.
The physical content of the models with intermediate symmetry containing
$SU(2)_L\otimes SU(2)_R$ and $D$ broken at the highest scale was first
stressed in \cite{chgi}. 

The ESH may be too drastic since in the 210 and 126 representations there
are multiplets with high quantum numbers, which may substantially
contribute to RGE. However, the mass spectrum of scalars depends on the
coefficients of the non trivial SO(10) invariants which appear in the
scalar potential, which can be constrained by requiring that the potential
absolute minimum is in the desired direction to give the considered
symmetry breaking pattern. This fact results in rather restrictive
conditions on scalar contributions to RGE. For details see last reference
quoted in \cite{bucc} or, for a summary of results, ref. \cite{buma}. 

\section{Phenomenology of SO(10) GUT's}

We discuss here three main phenomenological features of SO(10) GUT's and,
related to that, what experiments tell us on these models: proton
instability, neutrino masses and baryon asymmetry of the universe. In
particular we will only consider the models described in section 3. There
are actually others fascinating issues, as SO(10) inflationary models and
topological defect production, which however will not be covered here for
brevity (see on these topics for example \cite{esmi} and \cite{lama}). 

\subsection{Proton lifetime and $M_X$ GUT scale}

The stronger lower limit on proton lifetime comes from the channel $\pd$
\cite{gaje}, 
\be
\tau_\pd = \frac{\tau_p}{Br(\pd)} > 0.55\cdot 10^{33} years.
\ee
Further improvements on this value, as well as on partial mean lifetimes
for many other channels, are expected in the next few years from
Super-Kamiokande \cite{tots}. The ICARUS project \cite{rubb} should further
increase the present limits as well, in particular for exotic channels,
such as $p\rw e\,\nu\,\nu$, up to the range $10^{32}\div 10^{33}$ years.
The bound on $\tau_\pd$ can be translated into a lower limit on $M_X$,
which is the scale at which {\it leptoquarks} take mass \cite{bucc,leol}, 
\be
M_X = \left[ \frac{\tau_\pd}{10^{32}\, years} \right]^{\frac{1}{4}}~ 
10^{15}\,GeV \gapproxeq 1.5\cdot 10^{15}\,GeV.
\ee
From this lower bound one is therefore led to the conclusion that the two
models with D symmetry are ruled out (actually model B is at the very limit
of compatibility with experiments), while the ones based on $\tiz$ or $\lu$
intermediate symmetry are still in good shape, predicting partial lifetimes
for the $e^+\pi^0$ channel of the order of $9\cdot 10^{35}$ and $5\cdot
10^{34}$ years respectively. 

\subsection{Neutrinos and the $M_R$ breaking scale}

There are two experimental facts which are somehow suggesting that
neutrinos may be massive particles. On one hand the reduction of the
observed flux of solar neutrinos \cite{gall} with respect to the one
predicted by solar models \cite{bapi} may be explained in terms of MSW
neutrino oscillations \cite{wolf}. Furthermore evidences for DM in galactic
halos and at supercluster scales, together with studies on structure
formations, are in agreement with a massive neutrino with mass of few $eV$
\cite{prho}. 

The idea of neutrino oscillations was proposed long ago in pioneering works
by Bruno Pontecorvo \cite{pont} and has received a tremendous revival in
the last ten years, after it was realized that neutrino flux may undergo
resonant $\nu_e\leftrightarrow\nu_\mu$ or $\nu_e\leftrightarrow\nu_\tau$
transitions when passing through matter and in particular the solar
interior \cite{wolf}. This effect has been widely advocated as a solution
for the {\it solar neutrino problem} which we mentioned above. Actually
helioseismology \cite{ribe} constraints so much solar models that it
appears unlikely to reconcile the deficit in neutrino flux by an even
slight change in the sun central temperature. Furthermore the different
observed reductions for $^7Be$ and $^8B$ neutrinos are at variance with the
fact that both originate from the same parent $^7Be$ nuclei. Present
observations \cite{gall} require the following ranges for neutrino squared
mass difference and mixing angle (for oscillations of $\nu_e$ into a
$\nu_\mu$ or $\nu_\tau$): 
\be
\ba{lll}
\Delta m^2\sim 10^{-6}\div 10^{-5}\,eV^2, & sin^2 2\theta\sim 10^{-3}\div
10^{-2} & \mbox{(small angle solution)} \\ 
\Delta m^2\sim 10^{-5}\div 10^{-4}\,eV^2, & sin^2 2\theta\sim 0.2\div 1 & 
\mbox{(large angle solution)}. 
\ea
\ee

Before comparing these results with SO(10) GUT predictions let us also
briefly review what cosmological DM may tell us on neutrino masses.
Evidence for the existence of galactic DM was found as early as 1922 by J.
H. Jeans \cite{trim}. From observations on galactic rotation curves one
gets, for the actual to critical density parameter $\Omega = \rho/\rho_C$, 
\be
\Omega\geq 0.1,
\ee
while, looking at larger structures, cluster or superclusters \cite{dree},
\be
\Omega_{cluster}> 0.2\div 0.3.
\ee
Not all matter contributing to $\Omega$ is likely to be baryonic since the
baryon contribution $\Omega_b$ is strongly constrained by primordial
nucleosynthesis to be \cite{dree} 
\be
\Omega_b<0.1.
\ee

If massive, light neutrinos would contribute to $\Omega$ as \cite{cocl} 
\be
\Omega_\nu h^2 = \frac{m_\nu}{90\,eV}, \quad\quad\quad h\sim 0.54\div 0.73;
\ee
so $\nu_\mu$ and $\nu_\tau$ could easily give the inflation desired
prediction $\Omega=1$ without violating experimental limits on their masses
($m_{\nu_\mu} < 0.17\,MeV,~ m_{\nu_\tau} < 24\,MeV$ \cite{pdg}). However,
neutrinos are {\it hot} DM, i.e. they were relativistic when galaxy
formation started and structure formation models based on inflationary
schemes predict that {\it hot} DM only generates too few old galaxies
\cite{whfr}. Better agreement with data is obtained in case of mixed {\it
hot} + {\it cold} (non relativistic) scenario, with $\Omega_{hot\,
DM}\simeq 0.25$ and $\Omega_{cold\, DM}\simeq 0.7$. 

Going back to SO(10) GUT's, we already mentioned that light neutrinos are
predicted, in general, to get a mass via the {\it see-saw} mechanism
\cite{gera}, 
\be
m_{\nu_i} = \left( \frac{m_\tau}{m_b} \right)^2 \frac{m_{u_i}^2}{M_R} 
\frac{g_{2R}}{h_i},
\ee
where $u_i$ is the up quark of the same generation, $g_{2R}$ is the
$SU(2)_R$ coupling at the $M_R$ scale and $h_i$ the Yukawa couplings of the
$i$-th fermion generation to Higgs responsible for the symmetry breaking at
$M_R$. Using $m_\tau=1.777\,GeV$, $m_b=4.3\,GeV$, and $m_c=1.3\,GeV$, one
gets 
\bea
m_{\nu_\mu} &\sim& 2.9~ \frac{1}{M_R(10^{11}\,GeV)} \frac{g_{2R}}{h_2}~ 
10^{-3}\,eV, \nonumber \\
m_{\nu_\tau} &\sim& 5~ \frac{1}{M_R(10^{11}\,GeV)} \frac{g_{2R}}{h_3}~ 
10\,eV.
\eea
For the models C and D of section 3, which passed the proton lifetime test,
we therefore get 
\bea
\mbox{model C} &:& m_{\nu_\mu}\sim \frac{g_{2R}}{h_2}~ 10^{-3}\,eV,~ 
m_{\nu_\tau}\sim \frac{g_{2R}}{h_3}~ 17.5\,eV; \nonumber \\
\mbox{model D} &:& m_{\nu_\mu}\sim \frac{g_{2R}}{h_2}~ 4.3\cdot
10^{-2}\,eV,~ m_{\nu_\tau}\sim \frac{g_{2R}}{h_3}~ 750\,eV. 
\eea
If $g_{2R}/h_2$ is order of the unity, the model C with $\tiz$ intermediate
symmetry gives a value of $\Delta m^2=m_{\nu_\mu}^2 - m_{\nu_e}^2\sim
m_{\nu_\mu}^2$ of the right order of magnitude required by MSW solution to
{\it solar neutrino problem} with $\nu_e-\nu_\mu$ oscillation. In this case
$\nu_\tau$ would contribute to $\Omega$ with a fraction $\Omega_{\nu_\tau}
h^2\sim 0.2$, slightly larger than what desired in the {\it hot} + {\it
cold} scenario, which requires $m_{\nu_\tau}\sim$ 10 $eV$. The predictions
of model D seem less satisfactory: if the value for $m_{\nu_\mu}$ is still
in mild agreement with the large angle MSW solution (which is however {\it
theoretically} disfavoured) a too heavy $\nu_\tau$ is predicted, even
incompatible with the Cowsik and McClelland bound \cite{cocl}, $\sum_i~
m_{\nu_i}\lapproxeq 100\,eV$. It should also be mentioned that if one
releases ESH approximation one can get in the case of model D with $\lu$
intermediate symmetry the upper bound $M_R\lapproxeq 5\cdot 10^{11}\,GeV$
which turns into the lower bounds for light neutrino masses 
\be
\mbox{model D (without ESH):}\quad m_{\nu_\mu} \gapproxeq
\frac{g_{2R}}{h_2}~ 6\cdot 10^{-3}\,eV,~ m_{\nu_\tau} \gapproxeq
\frac{g_{2R}}{h_3}~ 100\,eV,
\ee
which are both at the boundary of values in agreement with MSW solution and
$\Omega_\nu<1$. 

\subsection{Baryogenesis via leptogenesis in SO(10) GUT's}

The value of $\Omega_b$ previously mentioned, obtained from the comparison
of experimental data on light nuclei abundances in the universe and
theoretical predictions on primordial nucleosynthesis, also gives another
important parameter, the present baryonic asymmetry normalized to photon
density \cite{dolg}, 
\be
\eta = \frac{n_B - n_{\bar B}}{n_\gamma} \simeq 3\cdot 10^{-10}.
\ee

Starting from big bang (likely) symmetric conditions $n_B=n_{\bar B}$, it
is clear that the value for $\eta$ requires at some stage of universe
evolution baryon number violating processes. Actually it was Sakharov who
pointed out thirty years ago the necessary conditions for the production of
a baryon asymmetry: 
\begin{enumerate}
\item
Baryon number violating interactions.
\item
C and CP violation.
\item
Non equilibrium conditions.
\end{enumerate}
It was soon realized that GUT's may be the natural framework for the
production of a finite value for $\eta$. In the standard scenario
\cite{kotu} it is generated by out of equilibrium decays of heavy Higgs or
gauge bosons. However, it was pointed out by several authors \cite{klma}
that anomalous B+L violating processes mediated by $SU(2)_L\otimes U(1)_Y$
sphaleronic configurations completely wash out any asymmetry produced at
GUT scales, unless a finite asymmetry is also present in the difference
B-L. This cannot be achieved in minimal SU(5) theory, for which B-L is a
global symmetry, but can be easily implemented in SO(10) GUT's, when
$U(1)_{B-L}$ is spontaneously broken at $M_R$. 

This possibility has been studied in ref. \cite{bumm} for the model D of
section 3, but a similar result for $\eta$ can be obtained for the {\it
favorite} case with $\tiz$ intermediate symmetry \cite{bump}. The mechanism
is based on out of equilibrium decays of Higgs bosons of the 210
representation into Majorana neutrinos at $M_R$. The resulting lepton
number asymmetry is then converted into baryon number at low scales via the
shuffling effects of sphalerons, giving a result for $\eta$ compatible with
the value required by nucleosynthesis. This baryogenesis via leptogenesis
scenario was first considered for heavy Majorana neutrino decays in
\cite{fuya}. 

There are actually several baryogenesis models (for a recent review see
ref. \cite{dolg}), but it is nevertheless worth pointing out the fact that
the production of an asymmetry in B-L is a rather natural and unavoidable
prediction of SO(10) GUT's. 

\section{Conclusions}

We have reviewed the status of a class of SO(10) GUT's, constraining the
predictions for the unification scales $M_R$ and $M_X$ with the most recent
available data on proton lifetime and SM gauge couplings. 

What we may conclude in short is that the two models with D parity seem to
be excluded since they predict a too low value for $\tau_\pd$ while the
ones with $\tiz$ (model C) and $\lu$ (model D) intermediate symmetry are in
this respect still in good shape. 

We have also pointed out that there is a quite huge low energy
phenomenology ranging from neutrino masses to cosmological observations of
light nuclei which may provide further and quite stringent constraints on
GUT's and are certainly providing a hint for what to look for beyond the
SM. 

By considering neutrino mass predictions within the {\it see-saw}
mechanism, we get rather intriguing values for $\nu_\mu$ and $\nu_\tau$ for
model C, 
\be
m_{\nu_\mu}\sim \frac{g_{2R}}{h_2}~ 10^{-3}\,eV,~ m_{\nu_\tau}\sim
\frac{g_{2R}}{h_3}~ 17.5\,eV, 
\ee
which, though the Yukawa couplings $h_i$ are unknown parameters, are of the
order of magnitude to fit in the MSW solution to {\it solar neutrino
problem} and to generously contribute to the {\it hot} component of dark
matter. Model D seems instead to provide a much larger value for
$m_{\nu_\tau}$ in the Extended Survival Hypothesis and so it is a bit
disfavoured, though it cannot presently be ruled out because of the poor
knowledge of SO(10) GUT Yukawa sector. For both models a prediction for
baryon asymmetry in agreement with the value known for $\eta$ seems {\it
unavoidable}. Next decade experiments will hopefully provide new
informations to either ruling out SO(10) GUT's or confirming their many low
energy and cosmological predictions.

\end{document}